# PROGRESS OF DIAMOND DIGITAL LOW LEVEL RF

P Gu†, C. Christou, P. Hamadyk, G. B. Christian, D. Spink and A. Tropp, Diamond Light Source, Oxfordshire, UK

*Abstract*

The first version of digital low level RF (DLLRF) for the Diamond Light Source storage ring and booster was developed with ALBA Synchrotron. Six systems have been built so far. Two of them are in routine operation controlling two normal conducting HOM-damped cavities in the Diamond storage ring. A third system is being used for cavity testing in the RF test facility (RFTF). The fourth system is being commissioned to control the second normal conducting booster cavity. The fifth DLLRF system is being prepared for the third normal conducting RF cavity in the storage ring.

A new DLLRF system based on SIS8300-KU with RTM has been developed and tested in the last few years. We are aiming to develop a common platform for the different RF systems in Diamond, including the storage ring, the booster and the Linac. It will also be our baseline design for the future Diamond II. Firmware, software and supporting hardware have been developed and tested. The Linac version with arbitrary waveform generator mode was tested successfully to generate flat top pulse from SLED in the high power test in the Linac. The storage ring version was also tested successfully in RFTF.

## INTRODUCTION

Diamond Light source has been providing beam for users since 2007. We have three major types of RF systems, the linac RF system, booster RF system and storage ring RF system. The linac RF system works at about 3GHz, while booster RF system and storage ring RF system work at around 500MHz. The booster operates with two 5-cell PETRA cavities. The storage ring RF system consists of two superconducting RF cavities and two normal conducting HOM-damped (NC) RF cavities, with a third normal conducting cavity being commissioned. A first version of DLLRF system has been developed together with ALBA in 2015. Nutaq Perseus 601X advanced mezzanine card (AMC) with Xilinx Virtex 6 FPGA was used as the core hardware. We have built 6 systems so far. Two systems have been operational with two NC RF cavities since 2019. The DLLRF system for the second booster cavity was installed and tested successfully with beam this year. The DLLRF system for the third NC cavity in the storage ring is being commissioned.

Following the successful experience of the first version of DLLRF and as better hardware became available on the market, we started developing our new DLLRF. The new design maximizes the common hardware components for all Diamond RF systems. Struck SIS8300-KU with rear transition module (RTM) was selected as the core hardware. RTM is the only difference between 3 GHz DLLRF (Linac) and 500 MHz DLLRF (booster and storage ring). We have designed and built the new DLLRF systems. The new DLLRF systems have been tested successfully in the Linac and in RFTF controlling a NC cavity. The Linac version has demonstrated flat-top output pulse from SLED RF pulse compressor. The new DLLRF is our baseline design for the RF systems of Diamond II, including DLLRF for harmonic RF cavity.

## DEPLOYMENT OF FIRST VERSION DLLRF

The description of our first version DLLRF system can be found in [1]. It was based on the Micro Telecommunications Computing Architecture (MicroTCA) standard, for its reliability, modularity and scalability. A commercial advanced mezzanine card (AMC), Perseus 601X with Virtex6 FPGA from Nutaq, is used as the core processor of the control algorithm. 16-Channel 14-bit ADCs and 8-channel 16-bit DACs on FPGA mezzanine cards (FMC) are used for analogue inputs and outputs.

Clock and local oscillator (LO) were generated locally from 500 MHz master oscillator (MO) in each system. RF signals are mixed with 479 MHz LO signal to generate 20.8 MHz IF signal. The IF signal is filtered before being sampled by ADC at 4 times the IF frequency. In the up-conversion chain, the DAC directly generates 20.8 MHz IF signal, which is mixed with LO, filtered and amplified. There is a PIN switch in the up-conversion unit that can be triggered by an interlock. A digital patch panel was also included to interface digital signals between the LLRF and other systems of the RF plants (tuner motor controllers, PLC controllers, MPS, etc).

Since its first test in the Diamond booster, the DLLRF system has been tested in the RFTF controlling NC RF cavity. Two NC RF cavities were installed in the storage ring in August and November 2017 in straights adjacent to the existing RF straight. Two DLLRF systems were then deployed for the control of two NC RF cavities in the storage ring in 2019. One installed DLLRF system is shown in Figure 1. The second 300 kW amplifier was split to feed the two NC RF cavities through 9-3/16 inch coaxial transmission line.

The control precision was measured with beam. The DLLRF can achieve 0.2% RMS in amplitude and 0.02 degree in phase. Phase noise of the closed loop between 10 kW and 60 kW, measured at a diagnostic point in the DLLRF front end, is similar to that of the oscillator used to drive the system with the addition of some noise in the 1 kHz to 10 kHz range. Spurs are visible at harmonics of mains frequency in both oscillator and DLLRF measurements, as shown in Figure 2.

---
† email address    pengda.gu@diamond.ac.uk

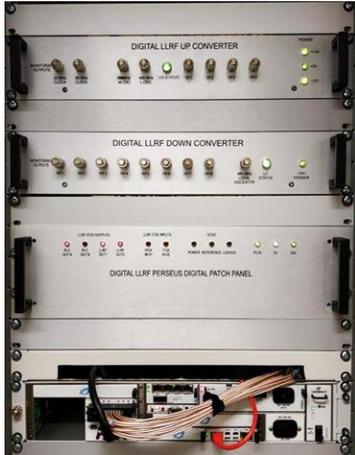

Figure 1. Installed DLLRF System.

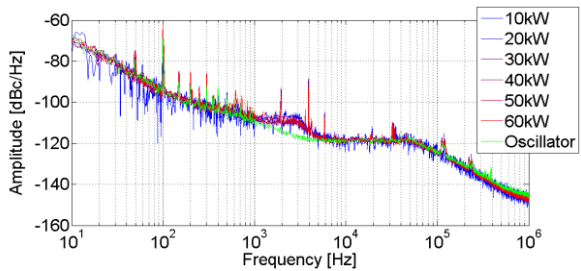

Figure 2: Measured Phase Noise

A second 5-cell Petra cavity was installed in the booster in 2018. A 60 kW solid state power amplifier (SSPA) from Ampegon was put into operation this year. DLLRF was quickly deployed for the conditioning and commissioning of the second booster cavity. The cavity was conditioned to over 50 kW in a week in June 2022. The electron beam from the Linac was captured and accelerated to 3 GeV using the second cavity. The waveform of cell 2 and cell 4 is shown in Figure 3 (a). Beam current in the booster is shown in Figure 3 (b).

The third NC RF cavity has been installed in the storage ring. A 120 kW SSPA from Cryoelectra has finished final acceptance test. Waveguide installation will be finished soon. DLLRF system is being installed.

The first version DLLRF worked reliably over the last few years. We had one Perseus 601X card failure due to broken DDR interface in 2019. It had to be sent back to Nutaq to be fixed. We also improved the firmware performance in the past few years. There were glitches while reading data out from FPGA. It was found to be a bug in timing constraints in the firmware. We also broke down the larger multiplexers for register read/write into pipelined stages by grouping the registers into banks of 16 to improve the reliability of the data transfer between the CPU AMC and the FPGA. The Mestor breakout box of Nutaq, used for digital patch panel, is now obsolete. We ordered 4 units as spares.

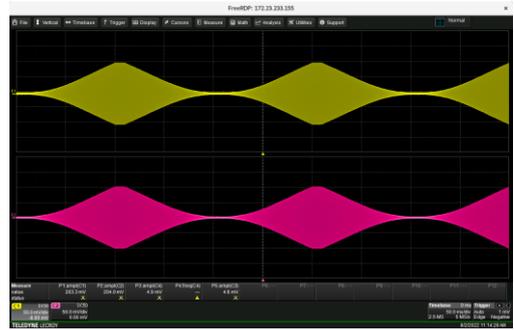

(a) RF Waveform in Cell 2 and Cell 4.

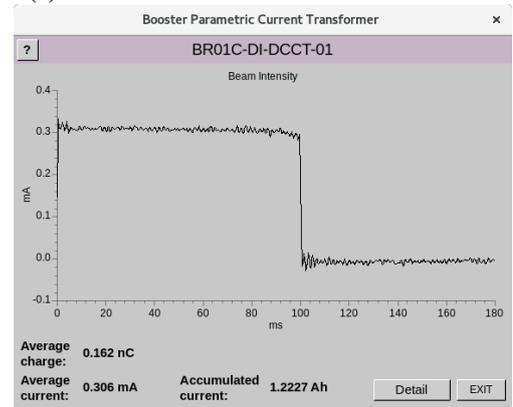

(b) Beam in the Booster.

Figure 3: DLLRF for Second Booster RF Cavity.

## NEW DLLRF DEVELOPMENT

The new DLLRF is also based on the Micro Telecommunications Computing Architecture (MicroTCA) standard. It consists of a 2U MTCA.4 chassis, an MCH, an AMC computer board, a Struck SIS8300-KU card and a Struck DWC8VM1 RTM with supporting clock/local oscillator (LO)/reference generation RF circuits. A comparison with the Nutaq system we used in the first version of DLLRF at Diamond is shown in Table 1.

Table 1. Comparison with previous DLLRF

|  | Struck System | Nutaq System |
| --- | --- | --- |
| FPGA | Xilinx Kintex Ultrascale | Xilinx Virtex 6 |
| Development Platform | Vivado | ISE & XPS |
| ADC | 10 ADC Channels, 125MS/s, 16-Bit | 16 ADC Channels, 125MS/s, 14-Bit |
| DAC | 2 DAC Channels, 250MS/s, 16-Bit | 8 DAC Channels 1000MS/s, 16-Bit |
| Interface to AMC computer board in MTCA | 4 Lane PCIe Gen3 | Ethernet |
| RF front ends | RTM in MTCA | External |
| Digital I/O | 4 inputs, 4 outputs | 16 inputs, 16 outputs |

*Hardware*

The DLLRF system architecture working at 500 MHz is shown in Fig. 4. DLLRF working at 3 GHz is slightly different, as shown in Figure 5. The two DLLRFs share many common components. They have the same MTCA, AMC computer board and SIS8300-KU. Their major differences are the RTM and Clock/LO generation. We will have a unified DLLRF platform for all three RF systems. This will greatly simplify the maintenance work in the future.

The AMC computer board is AM G64 from Concurrent. It is a double module with an Intel Xeon CPU and supporting PCIe Gen3. It communicates with the FPGA through PCIe bus. It will run Linux and EPICS. It will be the interface to the Diamond control system. NAT-MCH-PHYS MCH is used for the chassis management. The Struck SIS8300-KU has a Xilinx Kintex Ultrascale FPGA, 10 channels 16-bit ADC and 2 channels 16-bit DAC. It has a 4 lane PCIe Gen3 interface. The Struck DWC8VM1 has 8 channels downconverter and one vector modulator. IF signals are transferred through ZONE3 connector to FPGA. Struck cards are chosen for their proven performance in the European XFEL.

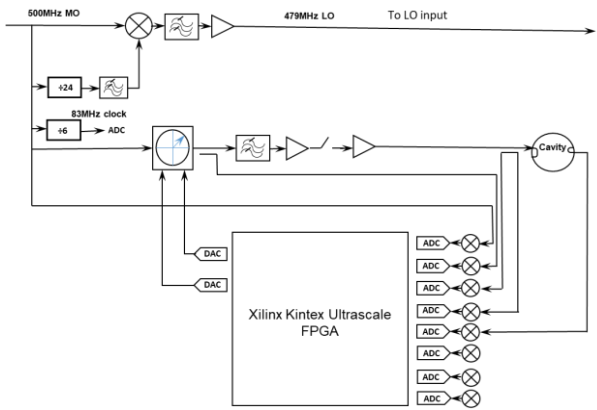

Figure 4: 500MHz DLLRF System Architecture.

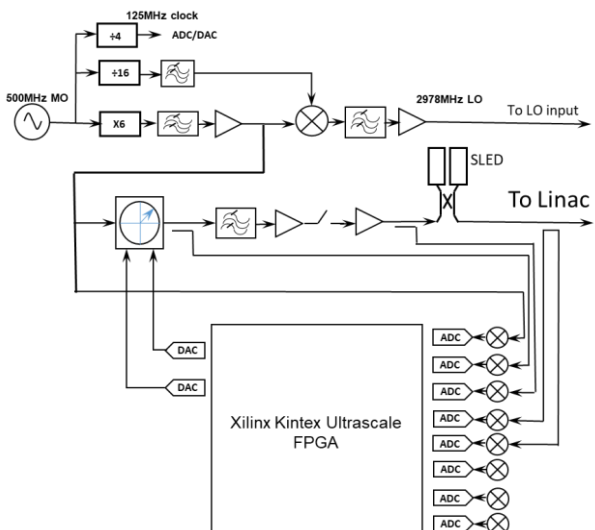

Figure 5: 3GHz DLLRF System Architecture.

**Clock and LO Generation** The structure of the clock and LO generation for 500 MHz DLLRF and 3 GHz DLLRF is very similar. Both of them use frequency dividers to generate IF and clock signal from the 500 MHz master oscillator. Clock frequency is four times the IF frequency, as IQ demodulation is used in the firmware. The 3 GHz DLLRF used a higher clock and IF frequency to maximize the sample number in the final 5 µs output pulse. The 3 GHz DLLRF also needs a frequency multiplier to generate the 3GHz signal from the master oscillator (MO). This method was chosen due to its simplicity. The number of components in the lock and LO generation were minimized to avoid introducing minimum additional phase noise.

The clock, IF, LO and 3 GHz are all derived from 500 MHz MO. They are naturally synchronized. The only problem is the phase uncertainty when there is a power down of the MO and the frequency divider. But it is limited to 4 options as I/Q demodulation is used. There is quadrant selection function implemented in the firmware. This uncertainty can be eliminated by comparing the phase measurement of MO signal before and after the power down. Another method that will be implemented in the future is the phase measurement will all be compared to the MO phase in the firmware. The second method has further advantage of reducing common phase measurement noise to all RF signals.

*Firmware*

The major modules of the firmware are shown in Fig. 6.

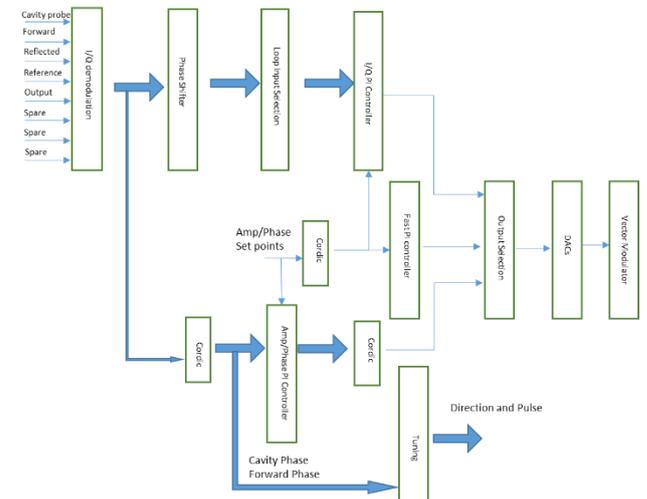

Figure 6: DLLRF firmware modules.

The 500 MHz DLLRF and 3 GHz DLLRF share many common modules, like IQ demodulation, Cordic, PI loop etc. The following functionalities are built into the FPGA firmware of 500 MHz DLLRF:

- IQ and polar PI loops of the cavity field to control amplitude and phase.
- Cavity tuning.
- Fast interlocks handling.
- Automatic start-up of the system.
- Automatic conditioning of the cavity
- Monitoring of RF signals.

The sampled IF signal is demultiplexed into IQ components. A counter controlled by the MO phase is used to control the demultiplexer. This enables the user to choose the sample to start with to establish the correct phase relation in the system after a power cycle. A software phase shifter has been implemented for each signal to keep the control loops stable. The firmware can be configured to run either in IQ PI loop or in polar PI loop mode. In the IQ loop mode, the phase set point and open loop measured phase should be in the same quadrant to keep the loop stable. In the polar loop mode, the CORDIC algorithm is used to translate IQ components to amplitude and phase before comparing them with the set points. The CORDIC algorithm is also used for the cavity tuner loop in 500 MHz DLLRF.

The 500 MHz DLLRF can be configured to run 3 modes, namely the normal conducting booster cavity mode, superconducting storage ring cavity mode and normal conducting storage ring cavity mode. In the booster cavity mode, cavity field ramping and tuning of the multi-cell cavity field flatness are also implemented.

**SLED Operation Requirement** A SLED RF pulse compressor has been installed and tested in Diamond Linac [2]. When operating in top-up mode for user beam the linac accelerates a single bunch of electrons, which is compatible with the generation of a peaked power pulse in the standard mode of SLED operation. To fill the ring, from empty, however, the linac must accelerate equally a train of up to 120 bunches. A standard phase switch mode operation results in around 7% energy spread and so a flat-top pulse is required. Following Benjamin Woolley and Igor Syratchev's paper [3], the coupling coefficient of SLED can be optimized. From the equation (15) (16) of the paper with some simplification, I and Q components of SLED input pulse should follow the equation 1) and equation 2) below.

$$F_x = \frac{R_x}{\alpha-1} + \alpha \left(1 - e^{-\frac{t1}{Tc}} - \frac{R_x}{\alpha-1}\right) e^{-\frac{(\alpha-1)(t-t1)}{Tc}} \quad 1)$$

$$F_y = \frac{R_y}{\alpha-1} - \frac{\alpha R_y}{\alpha-1} e^{-\frac{(\alpha-1)(t-t1)}{Tc}} \quad 2)$$

The amplitude and phase of the input and output pulse is shown in Figure 7. It can be seen clearly both phase and amplitude are modulated.

**3 GHz DLLRF** 3 GHz DLLRF works in pulse mode. Tuning module is not required in 3 GHz DLLRF. SLED operation requires the DLLRF to generate a pulse with phase and amplitude modulation. Phase and amplitude feedback will be needed to correct the pulse to pulse variation and long term drift. Phase and amplitude modulation is required in the final 1 μs. Three working modes were implemented in the firmware, namely standard mode with a simple phase switch, waveform (arbitrary waveform generation in the last 1 μs pulse) and full waveform mode (arbitrary waveform generation for the full 5 μs pulse).

There are two RJ45 connectors on the front panel of the SIS8300-KU, providing 4 LVDS digital inputs and 4 LVDS digital outputs. They are isolated and translated into TTL signals using CN0256 board from Analog Devices. These digital inputs and outputs are used for interlock signals.

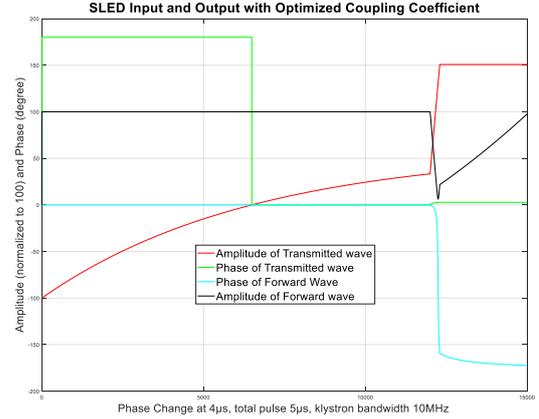

Figure 7: SLED flat-top pulse generation.

## NEW DLLRF TESTS

The new DLLRF systems were tested in the lab after the firmware and hardware were completed. All the functionalities were verified. The 500 MHz DLLRF was tested in the RFTF controlling a NC RF cavity. It showed similar behaviour as the first version. It can achieve better than 0.1% in amplitude and 0.03 degree in phase with all the control loops closed.

The 3 GHz DLLRF system went through several stages of tests. It first drove the SLED directly. The output waveform is shown in Figure 8. The flat-top output pulse was achieved using data calculated from equation 1) and 2) and corrected for klystron and modulator nonlinearities.

Next it was tested to drive pre-amplifier and the klystron to condition SLED to high power. Finally the output of SLED is connected to the Linac. Conditioning was quick. Electron beam was accelerated to 65 MeV with one klystron. The demodulated waveforms measured at the klystron output, pre-buncher, final buncher and accelerating structure 1 is shown in Figure 9. We reached around 2 times power gain from the SLED with a 1 μs flat top. This agrees to the design goal.

## CONCLUSION

The first version of Diamond DLLRF systems worked reliably in the last few years. New DLLRF systems have been developed for Diamond Light Source. The hardware, firmware and software have been tested successfully in RFTF and Linac. The new DLLRF system will provide a common platform for storage ring, booster and linac.

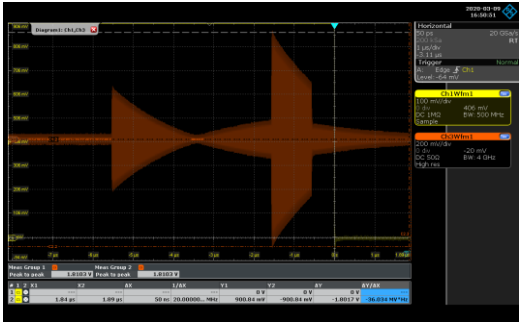

(a) Standard Mode

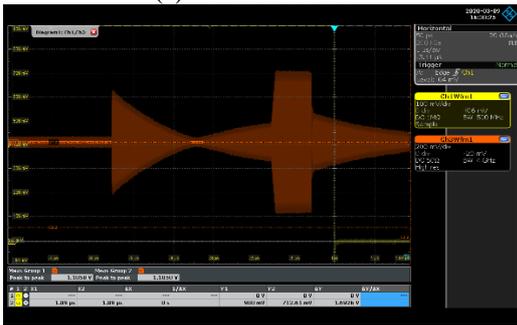

(b) Flat-top Output Pulse using Waveform Mode

Figure 8. 3GHz DLLRF Test with SLED

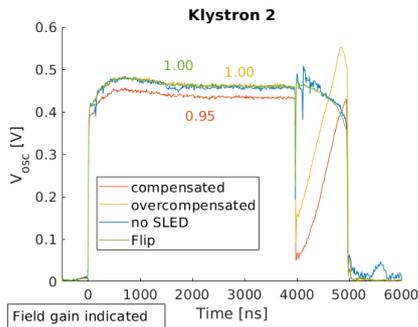

(a) Waveform from the Klystron

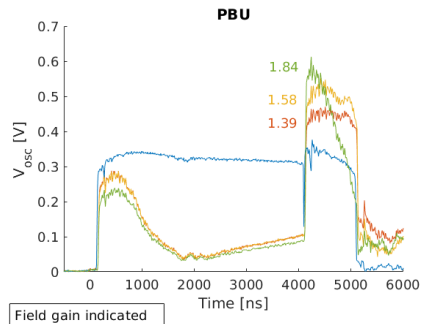

(b) Waveform at the Pre-buncher

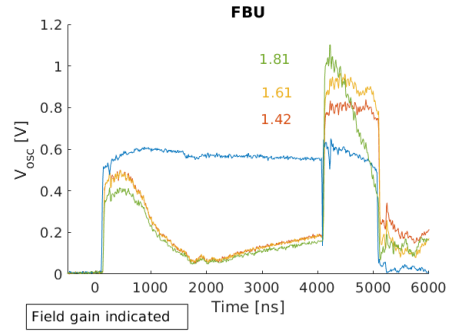

(c) Waveform at the Final Buncher

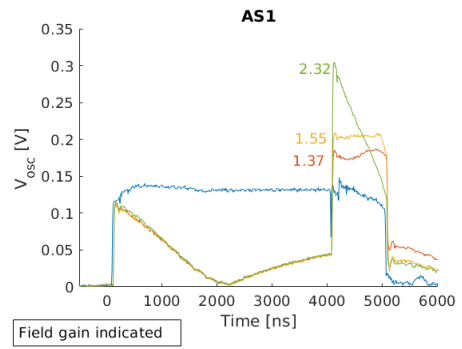

(d) Waveform at the Accelerating Structure

Figure 9: Demodulated Waveforms

## REFERENCES

[1] Pengda Gu, C. Christou et al., "Digital Low Level RF Systems for Diamond Light Source", in *Proc. 8th Int. Particle Accelerator Conf. (IPAC '17)*, Copenhagen, Denmark, May 2017, pp. 4089-4091.

[2] C. Christou, P. Gu and A. Tropp, "Programmable SLED system for single bunch and multibunch linac operation", in *31st Linear Accelerator Conference*, Liverpool, UK, August 2022.

[3] Benjamin Woolley, Igor Syratchev et al, "Control and performance improvements of a pulse compressor in use for testing accelerating structures at high power", in *PHYSICAL REVIEW ACCELERATORS AND BEAMS 20, 101001 (2017)*.